\documentclass[5p]{elsarticle}
\usepackage{lineno,hyperref}
\modulolinenumbers[5]
\usepackage[utf8]{inputenc}
\usepackage{float}
\usepackage{comment}
\usepackage{graphicx}
\usepackage[export]{adjustbox}
\usepackage{wrapfig}
\usepackage{lipsum}  
\usepackage{amsmath}
\usepackage{amssymb}
\usepackage{color}
\usepackage{mdframed}
\usepackage[normalem]{ulem}
\usepackage{hyperref}
\hypersetup{colorlinks=true, linktoc=all, linkcolor=blue, linktocpage, citecolor=blue}
\usepackage{color,soul}
\date{\today}

\def\NCC{N^{{\scriptscriptstyle(C)}}_{C}}
\def\NCP{N^{{\scriptscriptstyle(C)}}_{P}}
\def\NDP{N^{{\scriptscriptstyle(D)}}_{P}}
\def\NDC{N^{{\scriptscriptstyle(D)}}_{C}}
\def\NPP{N^{{\scriptscriptstyle(P)}}_{P}}
\def\NPC{N^{{\scriptscriptstyle(P)}}_{C}}
\def\NIJ{N^{{\scriptscriptstyle(i)}}_{j}}

\journal{Journal of Theoretical Biology}

\begin{document}
\begin{frontmatter}

\title{Symbiotic behaviour in the Public Goods Game with altruistic punishment}

\author[1]{Lucas S. Flores}
\author[1]{Heitor C. M. Fernandes}
\address[1]{Instituto de Física, Universidade Federal do Rio Grande do Sul, CP 15051, CEP 91501-970 Porto Alegre - RS, Brazil}
\fntext[corr]{Corresponding author: heitor.fernandes@ufrgs.br}
\author[2]{Marco A. Amaral}
\address[2]{Instituto de Humanidades, Artes e Ciências, Universidade Federal do Sul da Bahia, CEP 45638-000, Teixeira de Freitas - BA, Brazil.}
\author[1]{Mendeli H. Vainstein}
\fntext[corr]{Corresponding author: vainstein@if.ufrgs.br}

\begin{abstract}
Finding ways to overcome the temptation to exploit one another is still {a challenge in behavioural sciences}. In the framework of {evolutionary game theory}, punishing strategies are frequently used to promote cooperation in competitive environments. 
{Here, we introduce altruistic punishers in the spatial public goods game. This strategy acts as a cooperator in the absence of defectors, otherwise it will punish all defectors in their vicinity while bearing a cost to do so.}
We observe three distinct behaviours in our model: i) in the absence of punishers, cooperators {(who don't punish defectors)} are driven to extinction by defectors for most parameter values; ii) clusters of punishers thrive by sharing the punishment costs when these are low iii) for higher punishment costs, punishers, when alone, are subject to exploitation but in the presence of cooperators can form a symbiotic spatial structure that benefits both. 
This last observation is our main finding since neither cooperation nor punishment alone can survive the defector strategy in this parameter region and the specificity of the symbiotic spatial configuration shows that lattice topology plays a central role in sustaining cooperation. Results were obtained by means of  Monte Carlo simulations on a square lattice and subsequently confirmed by a {pairwise comparison of different strategies' payoffs in diverse group compositions,} leading to a phase diagram of the possible states.
\end{abstract}

\end{frontmatter}

\section{Introduction}

One of the most intriguing questions in evolutionary game theory is the emergence and maintenance of cooperation in an environment with limited resources and selfish individuals~\cite{Smith82, Nowak2006, Perc2017}. However, cooperation, in its broadest definition, is a phenomenon both frequent and extremely difficult to explain in terms of a mathematical model~\cite{Pennisi2005}.
Classical game theory predicts that unconditional betrayal should be the most rational choice in conflict situations~\cite{Szabo2007, nowak_n05}. Nevertheless, cooperation is commonplace not only among individuals of the same family, flock, or species but also in inter-species interactions ~\cite{Nowak2006, wilson_71,  Nowak2011a}, as in symbiosis or obligate mutualism. This phenomenon is very common in nature, such as in bacteria ~\cite{hosokawa2016obligate, yurtsev2016oscillatory}, animal gut microbiome ~\cite{Lewin-Epstein2020} and even regarding cells and mitochondria ~\cite{Henze2003}.  A memorable example of this situation happens with the acacia tree and ants of the species {\it Pseudomyrmex ferruginea}: the former depends upon the ants for the protection of its fruits against herbivores,  whereas the latter benefit by acquiring food and shelter~\cite{janzen1966coevolution}.

In the context of evolutionary game theory, a usual way to model the interaction among players is via the public goods game (PGG), in which groups are formed by agents that have two possible strategies, to cooperate or to defect ~\cite{Wardil2017}. Cooperators bear an individual cost to invest in a collective pool while defectors do not. The sum of all investments is subsequently multiplied by a positive factor that represents the synergistic factor of mutual investment. Then the total amount is equally divided among all players in the group, independent of their strategy. 
This dynamic leads to a clear temptation to defect, since one can gain at the expense of the rest of the group without bearing an individual loss. The most rational choice for an exploited cooperator would be to defect, avoiding to sustain players who do not contribute. Nevertheless, as all agents end up defecting, this results in a worse outcome than if all cooperated, a scenario called the Tragedy of the Commons~\cite{Szabo2007}.

Among many different applications, a current use of the PGG is the modeling of climate changes (collective risk dilemma) ~\cite{Yang2020, Curry2020, Gois2019, Vicens2018, Couto2020}, where it is possible to understand the consequences of the maintenance of cooperation and observe strategies to avoid catastrophic scenarios. 
Although humans are a highly cooperative species, historically we see that the exploration of natural resources, other species, and even of our own species generates deep impacts on the planet, due to the temptation of individuals and countries to increase their gain.
Several mechanisms were proposed with the aim of {explaining} cooperation in competitive systems, such as group selection~\cite{wilson_ds_an77}, spatial reciprocity~\cite{Nowak1992a, wardil_epl09, wardil_pre10, Zhao2020}, reputation~\cite{dos-santos_m_prsb11, brandt2003punishment}, direct and indirect reciprocity  ~\cite{nowak2006five, trivers_qrb71, axelrod_s81}, willingness ~\cite{brandt2006punishing, hauert2005game}, mobility ~\cite{vainstein2007does}, heterogeneity  ~\cite{perc_bs10, Amaral2016, Amaral2015, Amaral2020a, Amaral2020, Fang2019, Zhou2018}, among others such as the introduction of tolerance, multiple strategies and behavioural diversity~\cite{Wardil2017, Szolnoki2016a, Xu2017, Stewart2016, Hamilton1964, Junior2019}. 
 
When considering human interactions, a classic mechanism to promote cooperation is the punishment of defectors~\cite{Couto2020, yang2018promoting,boyd2003evolution, chen2014probabilistic, perc2015double, szolnoki2011phase, helbing2010evolutionary, rand2010anti, Fang2019a}. Such an approach consists of individuals who prefer
to withstand a small loss in order to harm defectors.
Based on these considerations, here we analyse the effect of altruistic punishers in the lattice PGG which besides contributing to the public pool like a normal cooperator also bear an extra individual cost in order to punish {each} nearby defector.

This strategy was proposed in previous works, showing the effects of heterogeneous punishment~\cite{perc2015double} and the counter-intuitive clustering of punishers with a protective layer of defectors~\cite{Szolnoki2017a}. It was also shown that,  independent of the cost, high punishment fines always favour punishers, who fight defectors more efficiently via segregation~\cite{Helbing_2010}. Nevertheless, cooperators are usually seen as second-order free riders in the parameter regions of the previous works. This generally leads to defectors being extinct by punishers allowing cooperators to dominate the system.

Here we explore parameter regions that allow punishers to take advantage of cooperators when the former cannot survive alone. This situation is the inverse of the second-order free-rider problem where {only} cooperators take advantage of punishers.
Specifically,  we focus on values that allow spatial structures of cooperators and punishers to resist defectors and grow in a symbiotic fashion.
We see that while well-mixed populations are unable to sustain such an arrangement, the topology of lattices allows specific formations that benefit both altruistic strategies in a scenario where none would survive alone. 

\section{Model}
We study a variation of the public goods game, in which three strategies are present: cooperators ($ C $), defectors ($ D $), and punishers ($ P $).
Players are located at the vertices of a square lattice of dimensions $ L \times L $, with periodic boundary conditions and interact with their first four neighbours (von Neumann neighbourhood). Initially, the strategies are randomly distributed with half of the players being $ D $, and the remaining half equally divided between $ C $ and $ P $, unless explicitly stated otherwise.
The interaction between players is such that cooperators and punishers contribute equally to a common pool with one unit (cost, $ c = 1 $), whereas defectors do not. The accumulated total in the group is then multiplied by a synergy factor ($ r>1 $) and divided equally among all players in the group, irrespective of their strategy.
When the player is a defector, a fine ($ \delta $) is discounted from its final payoff for each punisher present in its group. On the other hand, in addition to their investment, punishers also have to pay a fee ($ \gamma $) for each defector present in the group{, resulting in a decrease in payoff proportional to the number of defector neighbours.} We will use $ \gamma = \delta $ for simplicity.
Therefore, the payoff for a given agent is given by:
\begin{subequations} \label{eq.pay}
\begin{align}
\pi_C &= (r c/G)\,(N_{P}+N_{C}) - c \\
\pi_P &= (r c/G)\,(N_{P}+N_{C}) - c - \gamma\,N_{D} \\ 
\pi_D &= (r c/G)\,(N_{P}+N_{C}) - \gamma\,N_{P} ,
\end{align}
\end{subequations}
where $G$ is the group size, $N_C$ stands for the number of cooperative agents, $N_P$ for punishers and $N_D$ for defectors in the respective group, {including the central site}.

For each player's payoff, we only take into account the group in which the player is the central site. That is, differently from the typical PGG game, we consider that each player only receives gains from one group. The results are qualitatively the same if we consider that each agent belongs to five groups (data not shown), corroborating the findings of~\cite{szolnoki_pre09c,szolnoki_pre11c,hauert2003prisoner}.

The dynamics of the system evolves by means of a Monte Carlo simulation: first, a random site $ X $ is chosen along with a random neighbour $ Y $, and then both payoffs are determined and compared via an imitation rule. The agent in site $ X $  will adopt agent $ Y $'s strategy with a probability given by the Fermi function

  \begin{equation}
  \label{eq.transition}
     W_{X\rightarrow Y}= \frac{1}{1+\exp[-(\pi_Y-\pi_X)/K]} \,\,,   
 \end{equation}
where $ K $ is a noise associated with irrationality during interactions ~\cite{Szabo2007,szolnoki_pre09c}. 
A Monte Carlo step (MCS) is characterized by the above procedure repeated $L^2$ times, allowing all players to interact in a time step on average. In the simulations we used $K=0.1$, $G=5$, $ L = 100 $ and $ t_ {max} = 10 ^ 6 $ MCS. The simulation was stopped before $t_{max}$ if $D$'s became extinct, since we are interested in the proportion of altruistic strategies. We used averages over $100$ independent samples to generate  the results, unless  otherwise stated. 

\section{Results}

We begin by studying the general effects of the punishment parameter ($\gamma$) on a population.  It should be noted that if no defectors are present, punishers behave as normal cooperators; therefore, after the extinction of $D$'s,  both $C$'s and $P$'s are the same strategy, differing only by their labels, and any further dynamics will be the result of simple neutral drift~\cite{Nowak2006}. In other words, $C$'s and $P$'s will randomly fluctuate until one of them becomes the dominant strategy. Due to this, the sum of the cooperator and punisher densities, $\rho=\rho_C+\rho_P$, is more meaningful than the simple fraction of $C$'s or $P$'s in terms of equilibrium states. We refer to this quantity as the fraction of altruistic strategies throughout the paper.
 
\begin{figure}[t]
\begin{center}
\includegraphics[width=1.\columnwidth]{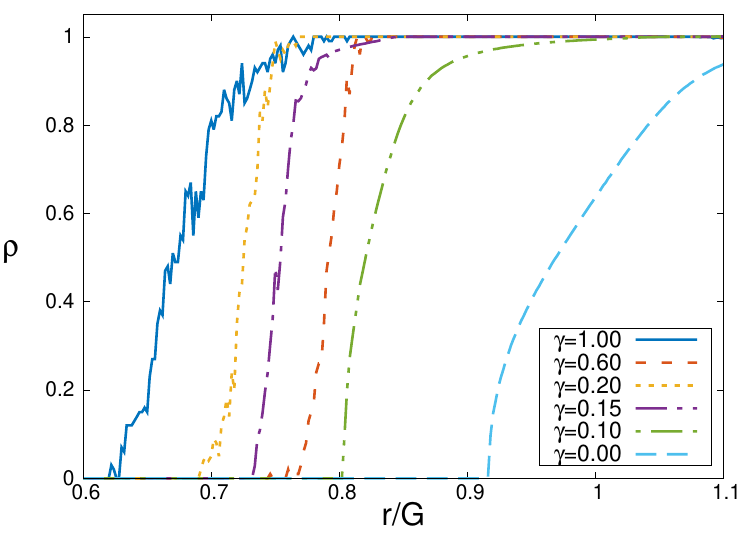}
\caption{Non-monotonic behaviour of the onset of cooperation as $\gamma$ increases. The graph shows the average final fraction of altruistic strategies, $\rho=\rho_C+\rho_P$, as a function of $r/G$ for different values of the punishment parameter $\gamma$. First, low fines ($\gamma<0.2)$ lead to an increase of the survival range of altruistic strategies. Then, intermediate fine values  ($0.2 < \gamma < 0.8$) can lead to a decrease in the survival range. Furthermore, for $\gamma>1$ we see altruism reemerging thanks to symbiosis between cooperators and punishers.}
\label{fig-varR}
\end{center}
\end{figure}

The main effect of the punishment factor is presented in Fig.~\ref{fig-varR}, that shows the average value of $\rho$ in the stable regime as a function of $r/G$. Note that for $\gamma=0$ punishers behave as cooperators, and we recover the classic PGG, in which cooperation is only able to survive if $r/G>1$ for the infinite well-mixed  population~\cite{szolnoki_pre11c}. 

\begin{figure}[t]
\begin{center}
\includegraphics[width=1.0\columnwidth]{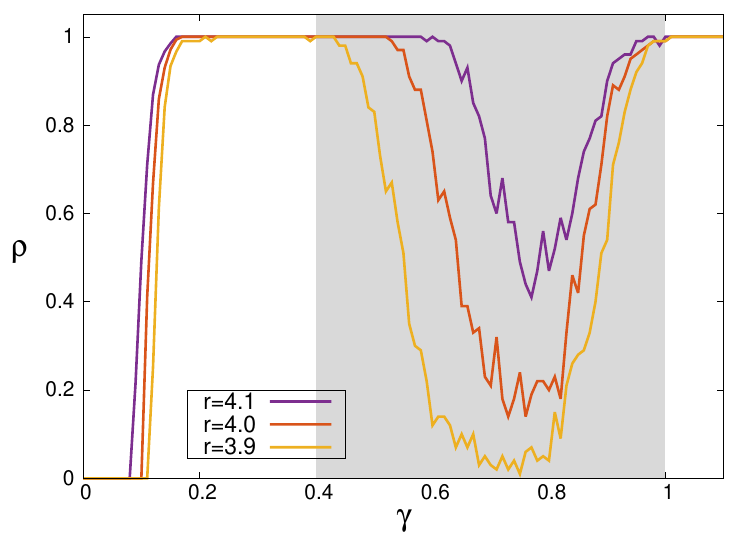}
\includegraphics[width=0.495\columnwidth]{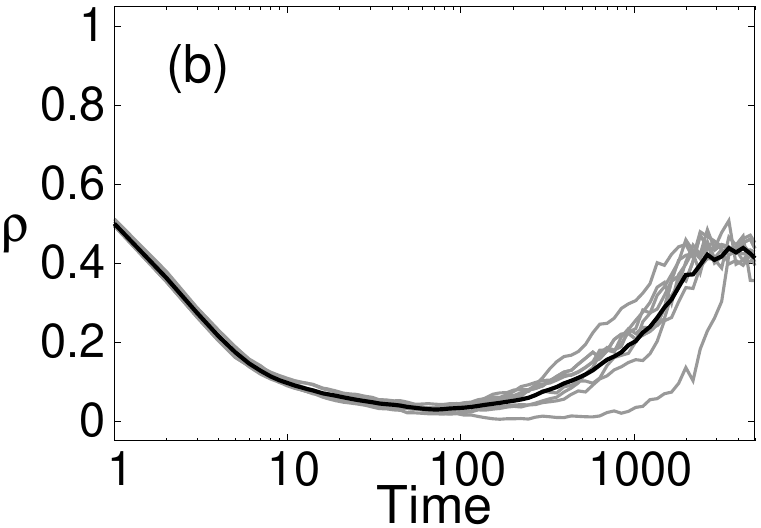}
\includegraphics[width=0.49\columnwidth]{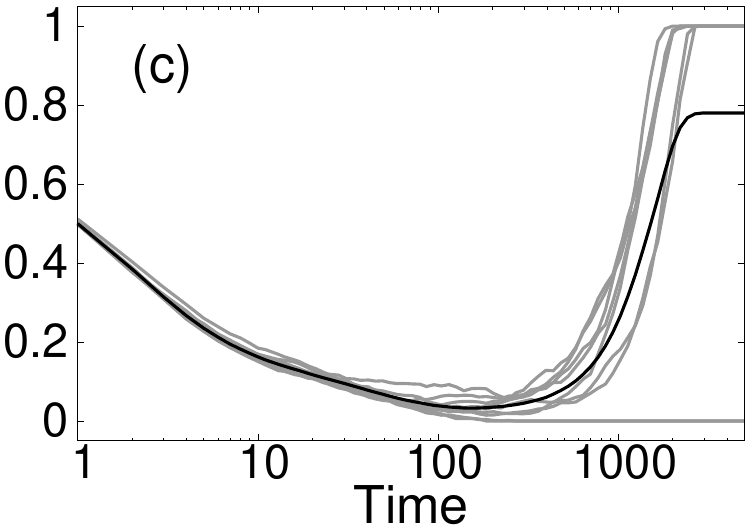}
\caption{ (a) Final fraction of altruistic strategies, $\rho=\rho_C+\rho_P$, as a function of $\gamma$. A small fine ($0.1<\gamma<0.4$) is sufficient to promote the extinction of defectors. For intermediate fines, highlighted by the grey shaded area, the system is lead to a bistable equilibrium regime, i.e., the lines represent the average of $\rho$, but either $\rho=0$ or $\rho=1$ for each sample. In the long run there is no coexistence of all three strategies. Panels (b) and (c) represent the temporal evolution of $\rho$. In (b), $\gamma=0.11$ and $r=4$, representing a region where the average behaviour reflects the sample behaviour. Note that all samples (light gray) fluctuate around the average value (black). For panel (c),  $\gamma=0.6$ and $r=4$, typical values for the  bistable region.}
\label{fig-vargamma}
\end{center}
\end{figure}

As expected, for low values of $r$ and $ \gamma $, both cooperators and punishers become extinct. In our model for $\gamma=0$ (that is, punishers behaving as cooperators), we see a continuous transition near $r/G\simeq0.9$, where cooperation begins to flourish. 
Two different phenomena are observed as the value of $\gamma$ is increased. First, the critical value $r^{*}$, where altruism starts to survive, changes in a nonlinear way. Note that as $\gamma$ increases, starting from zero, the extinction point first recedes in $r/G$, since the applied punishers' fines are able to act as a deterrent to defectors. In this region ($0<\gamma<0.2$), the fine value is low enough so that punishers can sustain the expense and,  at the same time, there are on average enough punishers around defectors to make the sum of the fines surpass the benefits of defection. 
However, this trend inverts for $\gamma>0.2${, where now the cost has a significant impact in the punisher payoff. From this point on,} further increases in the fine's value actually { impair punishers survivability, which in turn leads} to a larger $r^*$ necessary for the survival of altruism ($P$ and/or $C$). This effect is consistent until $\gamma \approx 0.8$, at which point altruistic strategies start to flourish again at lower values of $r^*$. This scenario suggests that punishment plays a nontrivial role on altruistic behaviour.
The second observed effect regards the qualitative change in the nature of transitions. We can observe that for small $\gamma$ values, the transition is continuous and smooth. Nevertheless, as $\gamma$ increases, the transition becomes steeper and the fluctuations increase drastically as the system switches to bistability, a behaviour that suggests a shift from a second-order-like phase transition to a first-order-like one. This is illustrated in Fig.~\ref{fig-vargamma}, where the effects of continuously varying $\gamma$ are presented for the combined fraction of punishers and cooperators for different values of $r$.
For relatively low punishments, $0.1<\gamma<0.2$, we see a sharp increase in $\rho$, which depending on the values of $r$, can lead to altruistic dominance.
Nevertheless, an increase in the punishment fine can also be detrimental to the altruistic strategies, as seen in the shaded region of the graph, corresponding to $0.4<\gamma<1$. Since the fine applied to defectors has the same value as the endowment paid by punishers,  there is inevitably a turning point where said costs become too large and the fraction of punishers (together with cooperators) starts to decrease again as we increase $\gamma$. 

We present the temporal evolution of several samples in Fig.~\ref{fig-vargamma} (b) for $\gamma=0.11$, $r=4$, representative values of the region in which the asymptotic value of $\rho$ fluctuates close to its average i.e. the system reaches a dynamic equilibrium. An important thing to note is that the model presents bistable equilibria in the shaded region of Fig.~\ref{fig-vargamma} (a), $0.4<\gamma<1$, In Fig.~\ref{fig-vargamma} (c) we present the temporal evolution for $\gamma=0.6,r=4$, where each sample evolves either to $\rho=0$ or $\rho=1$, and the average $\rho$ represents the fraction of samples that evolve to the fully cooperative state. 
Finally, for very high fine values ($\gamma>1$) we see that altruistic strategies dominate the system again. We explain the mechanisms underlying this change in behaviour in the following section.

\subsection{Phase Diagram}
\label{sec.phasediagram}

\begin{figure*}
\includegraphics[width=0.38\textwidth]{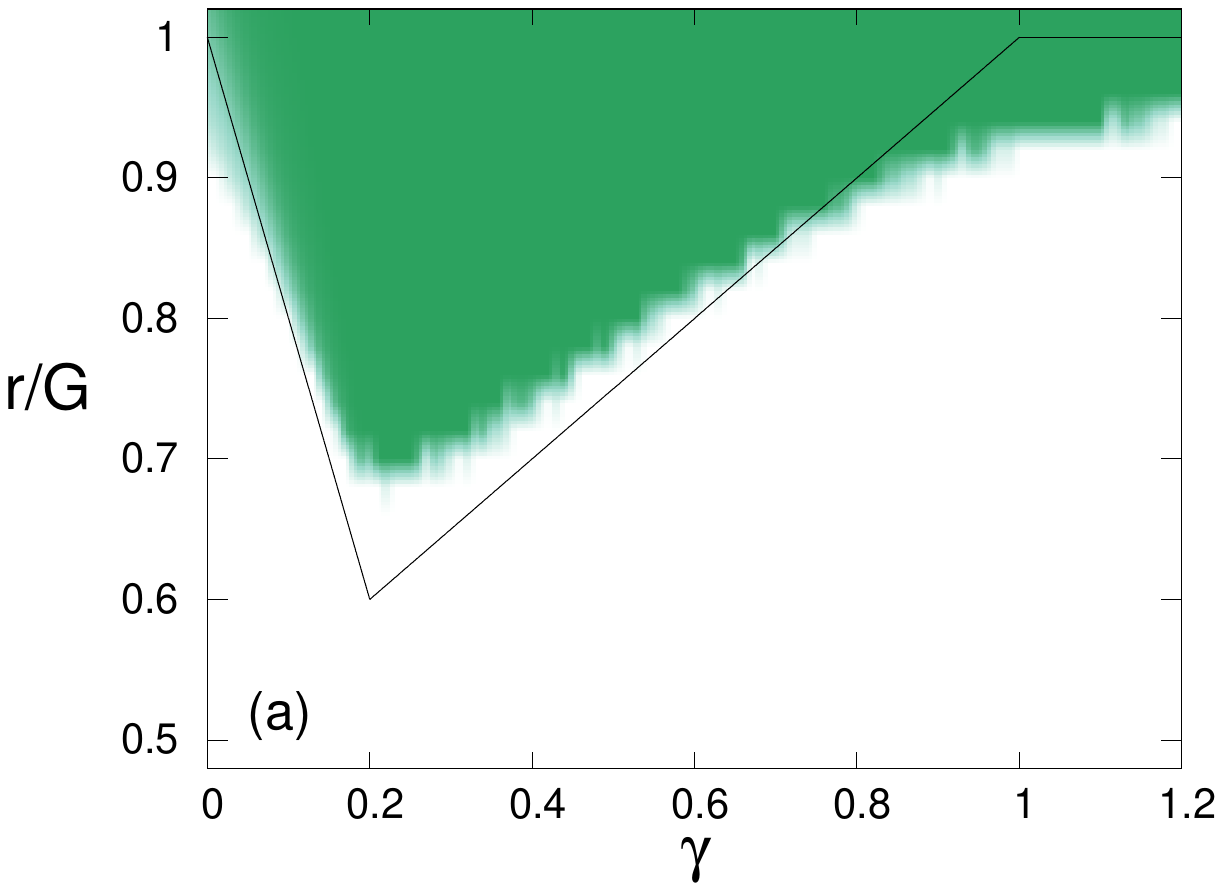}\hspace{-1.5cm}
\includegraphics[width=0.38\textwidth]{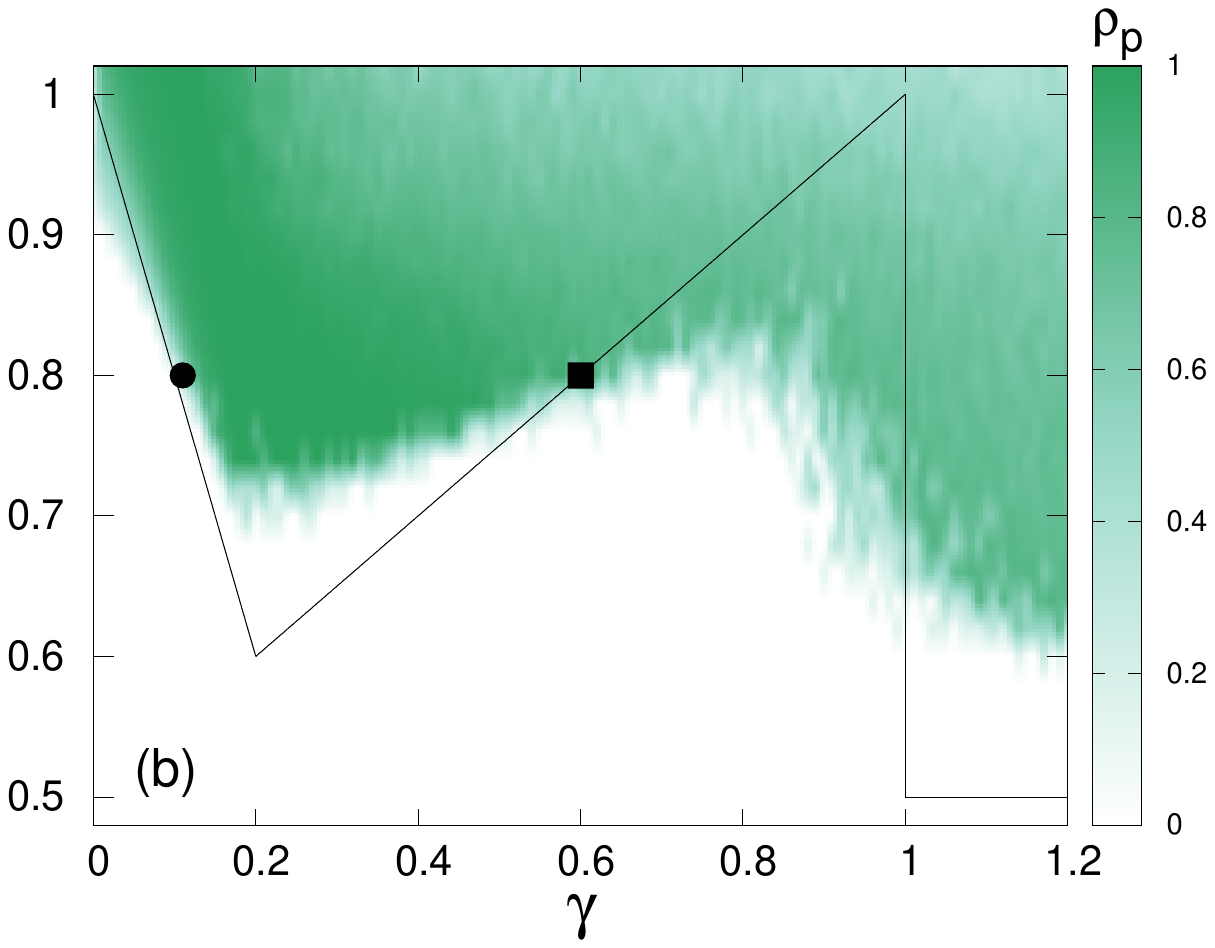}\hspace{-1.5cm}
\includegraphics[width=0.38\textwidth]{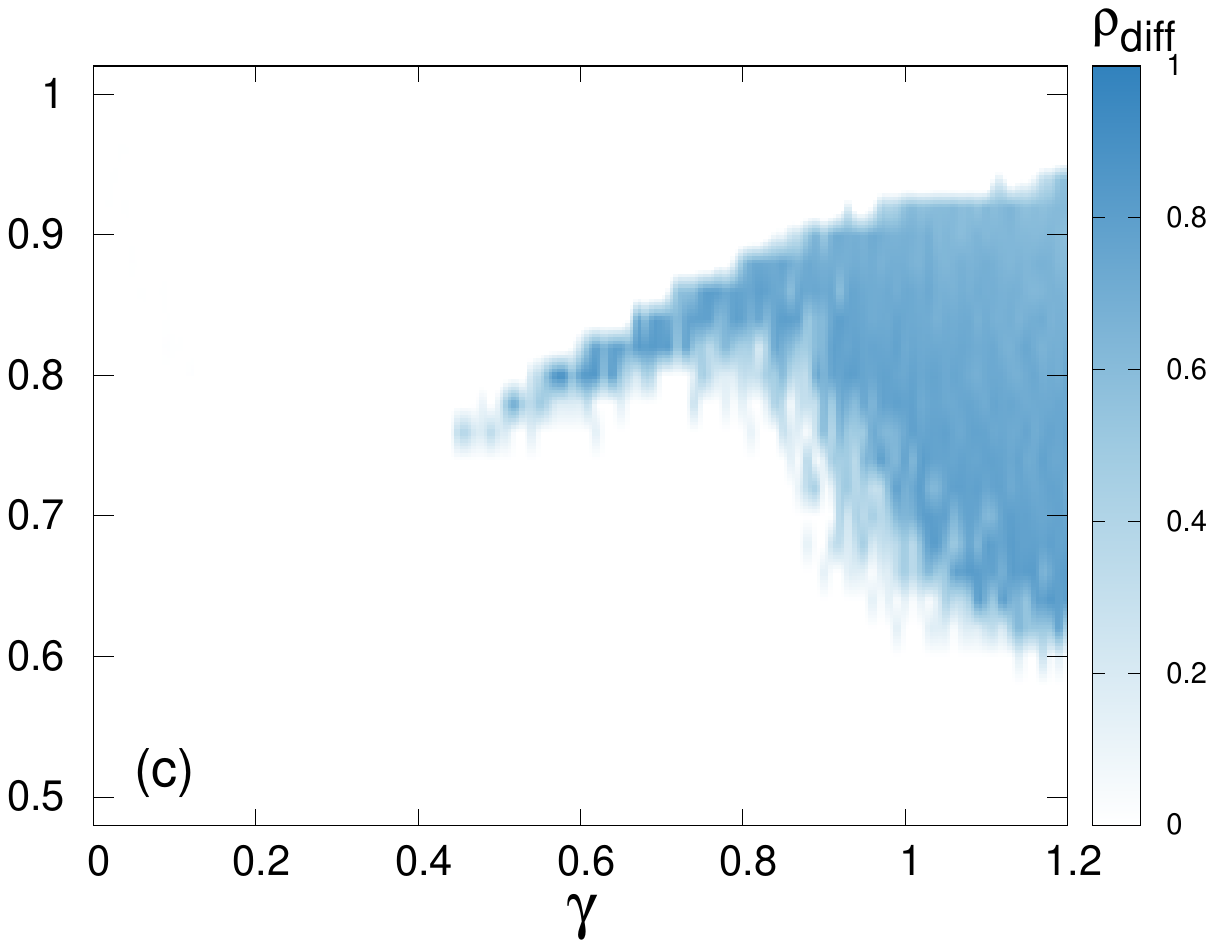}
\vspace{-0.7cm}
\caption{Parameter diagram, $ r/G \times \gamma $ for the noise parameter $K=0.1$; the color gradient represents the fraction of punishers in panels (a) and (b). In panel (a) the system is composed of only punishers and defectors (cooperators are absent). Here, $D$s dominate the population for low $r$ values in general. The major exception is the V-shaped region ($0<\gamma<1$),  where punishers are able to form compact clusters that survive defectors by sharing the fine cost among themselves (see main text for details). As $\gamma$ increases, this becomes harder and demands a higher $r$ for the survival of $P$s. Notice that on the vertical axis, where $\gamma = 0$, $P$s are equivalent to cooperators.  Panel (b) shows the situation in which all three strategies are present.
{The circle and square symbols corresponds respectively to the parameters used in Figure} ~\ref{fig-vargamma} {(b) and (c) for the stable and bi-stable regions.}
The striking difference between panels (a) and (b) regards the region of high $\gamma$: while neither cooperators nor punishers alone are able to survive for $r/G<0.9$ and $\gamma>1$, together they suppress defectors. This is the symbiosis region, where both $C$s and $P$s  arrange themselves in specific spatial patterns to survive. The continuous black lines in the first two panels represent the transition lines obtained from {the comparison of payoffs of two focal sites with different strategies (see appendix)}. As the noise parameter $K$ decreases, the simulation transitions approach the analytical results. 
Panel (c) presents the limits of the symbiosis domain, i.e.,  the region where the difference between the punisher fractions of panels (b) and (a), $\rho_{diff}$, is positive.
}
\label{fig.par-space} 
\end{figure*}

We summarize the general effects of varying the main model parameters ($ r $ and $ \gamma $) in the phase diagrams displayed in Fig.~\ref{fig.par-space}.
In Fig.~\ref{fig.par-space}(a) we present the final fraction of punishers in a specially prepared scenario composed only of punishers and defectors initially distributed equally (cooperators are absent). The first thing to notice is that while high $r$ can lead to punishers dominating, they can only survive with low $r$ ($r/G<0.9$) for $\gamma<1.0$. In this V-shaped region, the dynamics of the system is slightly dependent on $r$ and $\gamma$ but the outcome is the extinction of defectors except for the two transition regions: for the first one, with low gamma values there may be coexistence, and for the second, bistability. Punishers are able to survive since they mitigate the detrimental effect of the applied fines by forming stable clusters that divide the cost among the members. Near the first transition region, punishers and defectors coexist even for $r$ as low as $r/G \approx 0.7$. 
Close to the second transition region  ($0.2<\gamma<1.0$),  defectors invade all small clusters of punishers present in the initial configuration and can drive punishers to extinction, unless at least one of the remaining clusters reaches a critical size, which then expands until it dominates the entire population.

At high $\gamma$ values, altruism inevitably became unsustainable, since the punisher's payoff becomes too small  due to the high applied fines, and the clusters are unable to sustain themselves. Without cooperators, punishers are quickly eliminated for high $\gamma$ ($\gamma>1.0$). 
Note that in the case of a population composed only of defectors and cooperators ($\gamma = 0$), cooperation also becomes extinct for low $r$ ($r/G < 0.9$, see Fig.~\ref{fig-varR}). In other words, the region of high $\gamma$ and low $r$ is forbidden both for cooperators and punishers when they exist by themselves in a sea of defectors.

 In Fig.~\ref{fig.par-space} (b), we show the fraction of punishers for the initial configurations where all three strategies are present. Here, the regions of the parameter space dominated by defectors show a distinct behaviour as $r$ and $\gamma$ increase. While an increase in $r$ always leads to the eventual extinction of defectors, this is not the case for an increase in $\gamma$. For low values of $r$ ($r/G<0.6$), the punishment cost has no real effect on the final population outcome since defectors always dominate. However, as we increase $r$, we see that there are specific ranges of $\gamma$ that can lead to the emergence of altruism in a nontrivial way.

Comparing Figs.~\ref{fig.par-space} (a) and (b), we note that the low-$\gamma$ ($\gamma<0.2)$ region is essentially the same in both figures. The reason is that after the initial formation of clusters of cooperators and punishers, the former are spontaneously segregated to the boundaries, where they are exploited by defectors (they cannot survive alone for $r/G<0.9$). At intermediate $\gamma$ ($0.2<\gamma<1.0$), clusters of cooperators and punishers with a critical size can expand and dominate the system. After that, one of them will prevail due to neutral drift (usually punishers, since they compose a higher fraction of the population at the onset of defector extinction). As $\gamma$ increases, it gets more difficult for punishers to thrive by themselves, therefore cooperators persist for a longer time.

This situation continues until $\gamma$ reaches high values. Naively, we could expect that once $\gamma>1.0$, punishers would not be able to expand by themselves anymore, vanishing together with cooperators. However, looking at the phase diagram, Fig.~\ref{fig.par-space}(b), we note that this is not the case. In reality, we observe that in this region both cooperators and punishers coexist. We designate this the symbiosis region, a combination of parameters that allow both strategies to thrive where none would be able to do so by themselves. 

Finally in Fig.~\ref{fig.par-space}(c), we present the difference, when positive, between the fraction of punishers in the presence of cooperators, Fig.~\ref{fig.par-space}(b), and when alone, Fig.~\ref{fig.par-space}(a) to highlight the symbiosis region.
By doing so, we can clearly visualise the parameter region that allows the existence of punishers when cooperators are present, as opposed to the scenario with only punishers and defectors. We observe that the symbiosis region emerges as soon as $\gamma>0.5$, but only for a very small range of $r$ values. For $\gamma>1$ the symbiosis region becomes wider ($0.6<r/G<0.95$) and altruistic strategies are able to survive together.

\subsection{Emergence of Symbiosis}
\label{sec.symbiosis}

\begin{figure}[t]
\begin{center}
\includegraphics[width=0.42\columnwidth]{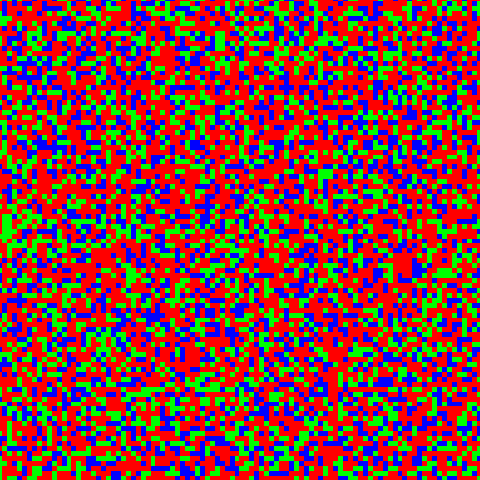}
\includegraphics[width=0.42\columnwidth]{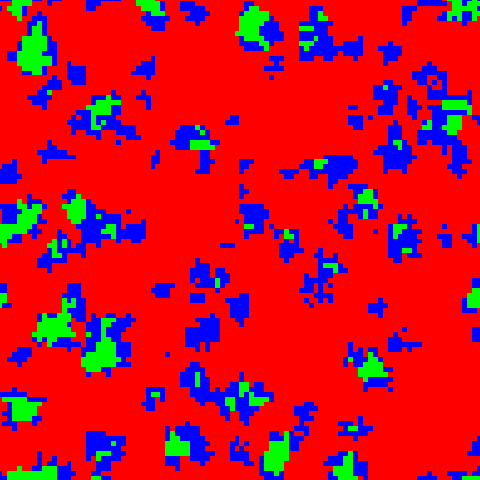} \\
\vspace{0.1cm} 
\hspace{0.05cm} 
\includegraphics[width=0.42\columnwidth]{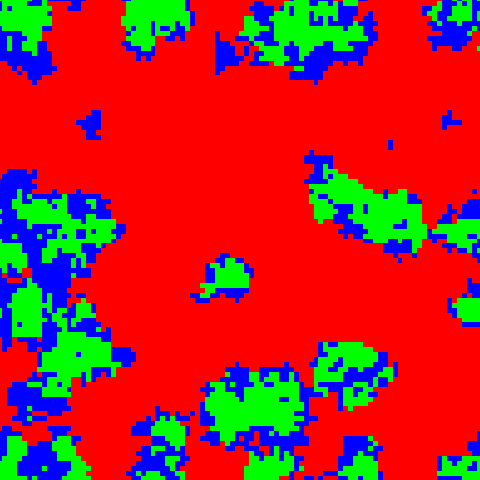}
\includegraphics[width=0.42\columnwidth]{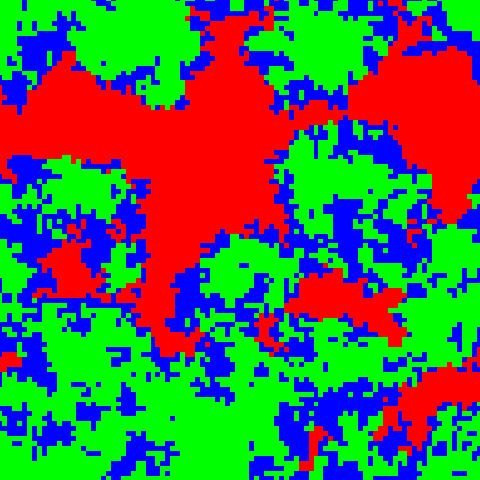}
\caption{Lattice representative snapshots showing the temporal evolution of the strategies for $r = 3.8$, $\gamma=1.5$ {and $K=0.1$}. Here the initial condition is a random strategy assignment with half of the sites being defectors (red, light grey), and the remaining half equally distributed between cooperators (blue, dark grey) and  punishers (green, lightest grey). From left to right and top to bottom, the snapshots present the Monte Carlo steps $t=0, 25, 130$ and $310$. For very early times, the cooperator and punisher densities drop, as unfavourable configurations are eliminated. Nevertheless, the symbiotic clusters with a bulk of punishers surrounded by a layer of cooperators survive this initial period and quickly begin to grow, dominating the entire population. In ~\cite{gif_fig4} we present a short animation of the lattice evolution.}
\label{figtimeevo2}
\end{center}
\end{figure}

Now, we focus on the region where the symbiosis effect occurs, see Fig.~\ref{fig.par-space}(c). To better understand the underlying mechanism responsible for this effect, we { analyse the payoffs of different strategies under diverse group configurations to characterize transitions, which could occur when these payoffs become equal}~\cite{Vainstein2014b}.{ To this end, we compare two neighbouring focal players of different strategies, each with its own $G=5$ agent group, to determine the parameter values that lead to the dominance of one strategy over another.}
{For instance, if we wish to compare a punisher player's payoff with that of a defector, first we  use } Eqs.~(\ref{eq.pay}) {to obtain}
\begin{subequations}\label{eq.payPD}
\begin{align}
\pi_P &= (r c/G)\,(N^{{\scriptscriptstyle(P)}}_{P}+N^{{\scriptscriptstyle(P)}}_{C}) - c - \gamma\,N^{{\scriptscriptstyle(P)}}_{D} \\ 
\pi_D &= (r c/G)\,(N^{{\scriptscriptstyle(D)}}_{P}+N^{{\scriptscriptstyle(D)}}_{C}) - \gamma\,N^{{\scriptscriptstyle(D)}}_{P},
\end{align}
\end{subequations}
{where $\NIJ$ represents the number of sites with strategy  $j$ in central agent $(i)$'s  group, including itself (e.g., $N^{{\scriptscriptstyle(D)}}_{P}$ is the number of  punishers in the group where the defector is the central site). Then we equate both payoffs in Eqs.}~(\ref{eq.payPD}){ to obtain all possible dominance transitions between strategies. By doing so, we can obtain the precise value of $r$ that makes both payoffs equal, let us call it $r_{PD}$  for the comparison between a punisher and a defector. To simplify the notation we write $r_{PD}\equiv r_{PD}(\gamma, \NPC+\NPP,\NDC,\NDP) $, given by}
\begin{equation} \label{eq.payPD-result}
r_{PD}=\frac{G}{c} \; \frac{\gamma[\NDP+(\NPC+\NPP)-G]-c}{\NDP+\NDC-(\NPC+\NPP)}
\end{equation}
{which is a function of $\gamma$ for each microscopic configuration encoded in the $N^{{\scriptscriptstyle(i)}}_{j}$'s, with $\NDP+\NDC \neq \NPC+\NPP$. It should be noted that there are as many transition values as there are configurations with the simultaneous constraints $\NDP\geq 1 $, $\NPC+\NPP \leq 4$,  and $N^{\scriptscriptstyle(i)}_C + N^{\scriptscriptstyle(i)}_D + N^{\scriptscriptstyle(i)}_P=G$ for $i\in\{P,D\}$;  however, not all of these possible transition values obtained in this way are relevant for the evolution of the system (see Appendix).}
The same procedure is repeated with a focal cooperative site and its defector neighbour. In this case, we obtain the {function}  $ r_{CD} \equiv r_{CD}(\gamma, \NCC+\NCP,\NDC,\NDP) $ {, given by}
\begin{equation} \label{eq.payCD}
r_{CD}=\frac{G}{c} \; \frac{\gamma \NDP-c} {\NDP+\NDC-(\NCC+\NCP)} \;,
\end{equation}
with $\NDP+\NDC\neq \NCC+\NCP$ {and the constrains $\NDC \geq 1 $, $\NCC+\NCP \leq 4$, and $N^{\scriptscriptstyle(i)}_C + N^{\scriptscriptstyle(i)}_D + N^{\scriptscriptstyle(i)}_P=G$ for $i\in\{C,D\}$. However, the transition values arising from this condition coincide with the constant and decreasing lines  obtained from }  Eq.~(\ref{eq.payPD-result}).  {The exceptions occur when} $\NDP+\NDC-(\NCC+\NCP) = 0$, {so that when we make $\pi_C=\pi_D$, we arrive at the vertical transition lines}
\begin{equation}
\label{eq.vertical-transitions}
\gamma_{CD} =\frac{c}{N^{\scriptscriptstyle(D)}_P},
\end{equation}
where $ c = 1 $. {Out of these, the only important line is $ \gamma_{CD} = 1 $, which corresponds to $N^{\scriptscriptstyle(D)}_P=1$ (see Appendix).}
{The relevance of this vertical transition line is that}  for high enough fines ($ \gamma \geq 1 $), a cooperator can invade a defector that is being currently punished. This highlights the importance of the interdependence between cooperators and punishers in order to defeat a defector.
We proceed with our analysis looking at representative snapshots of the lattice evolution in the symbiosis region in  Fig.~\ref{figtimeevo2}, where we observe a very interesting phenomenon: cooperators and punishers cluster in specific patterns.
Differently from other parameter regions, we see that {here} $C$ and $P$ strategies tend to form compact clusters in which cooperators lie in a thin boundary layer around a bulk composed of punishers.
Note that due to the inherent random nature of the simulations, {such} configurations are susceptible to noise, although their general shapes are frequent and occur for many different cluster sizes.
Since the initial condition is random, there will be some clusters that combine punishers and cooperators in such a way that they can survive the initial invasion of defectors. These symbiotic configurations will grow pushing cooperators to the boundaries and leaving punishers inside. At the same time, spatial configurations that do not favour symbiosis will be quickly eliminated by defectors. 
In~\cite{gif_fig4} we present a short animation of the lattice evolution, where it is possible to see such processes. 
Since only cooperators and punishers contribute to the common pool, low $r$ values have a stronger impact on their payoff than on that of defectors. Consequently, this symbiotic behaviour disappears for low $r$, leaving only defectors.

\begin{figure}[t]
\begin{center}
    \includegraphics[width=1.0\columnwidth]{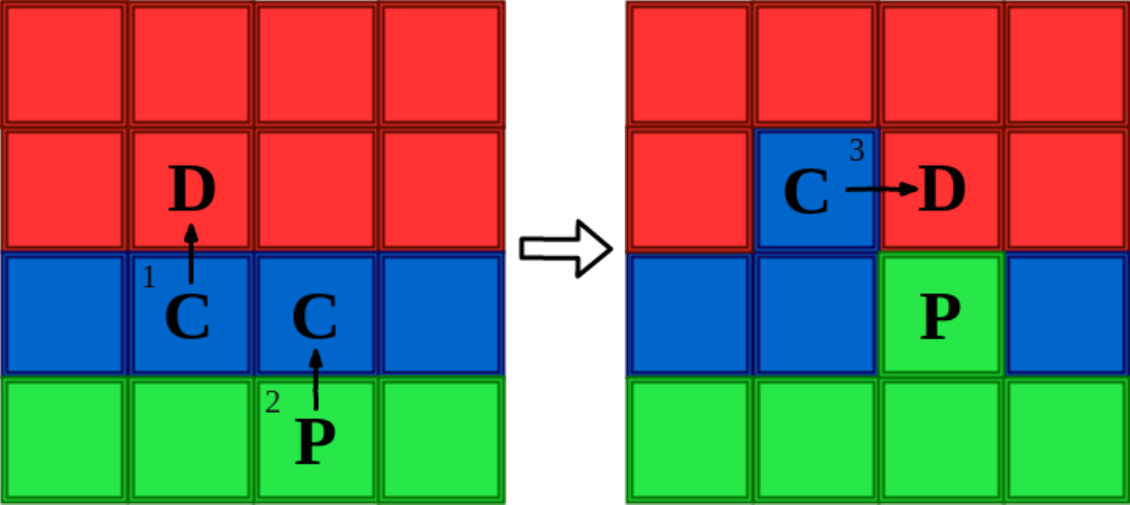}
\caption{ Illustrative diagram of the microscopic spatial configuration that characterizes symbiosis and allows $C$ to invade $D$ with the aid of $P$. In step 1, a $D$ will be invaded by $C$ for  $r/G > 1/3$. In step 2, $P$s are able to invade $C$s located at the borders in any scenario, due to contact of $C$s with $D$s. The symbiosis effect happens in step 3 during which the $C$ site would have a lower payoff than any of its $D$ neighbours in the absence of $P$s. However, in the presence of $P$s and for high values of punishment, $\gamma > 1$,  the defector's payoff is decreased allowing the invasion by $C$. The condition $r/G > 0.5$ (horizontal line in phase diagram, Fig.~\ref{fig.par-space} (b)) prevents the $D$ from invading the neighbouring $P$ in the right panel.}
\label{desenho}
\end{center}
\end{figure}

The spatial structures that promote symbiosis can be better understood by looking at the microscopic interactions when there is a cluster of punishers surrounded by a thin layer of cooperators that separate them from defectors. Fig.~\ref{desenho} illustrates this spatial configuration and all steps necessary for altruistic behaviour to thrive.
The first step is the expansion of the cluster as a cooperator invades a defector in the sea of defectors. This initial configuration does not favour the cooperator (for low values of $r/G$) because, now, the cooperator has several defectors as neighbours. However, the second step is the invasion of a boundary cooperator by a punisher from the inner cluster. This step has a high probability of happening since boundary cooperators are exploited by defectors and inner punishers are not. 
At this point, the condition $r/G > 0.5$ prevents the $D$ from invading the neighbouring $P$ site.
Once both movements occur and a {the system arrives at a } configuration such as the right panel of Fig.~\ref{desenho}, the comparison of payoffs between the cooperator and the defector shows that for $\gamma>1$ the cooperator will be able to invade the defector, {according to}  Eq.~(\ref{eq.vertical-transitions}). After this invasion succeeds, the system is, again, in a configuration where all these steps can repeat. 

It should be pointed out that the punisher alone cannot expand due to the high punishment cost associated with invading a sea of defectors.
This is the main mechanism that allows the symbiosis of cooperators and punishers. We can see that this is a complex dynamics where cooperators and punishers both play specific roles regarding cluster growth. 
We stress that this is a nontrivial scenario that is only possible in a specific range of parameters.
There is a delicate equilibrium between all involved factors, that is, the punishment fines and cost, the number of neighbours in a given configuration, the maximum allowed number of neighbours in the current lattice topology, etc.

\begin{figure}[t]
\begin{center}
\includegraphics[width=.42\columnwidth]{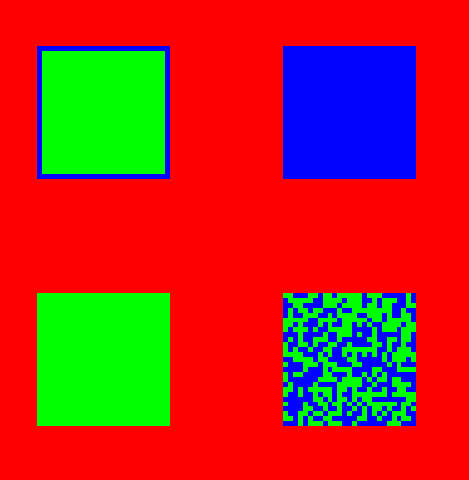}
\includegraphics[width=.42\columnwidth]{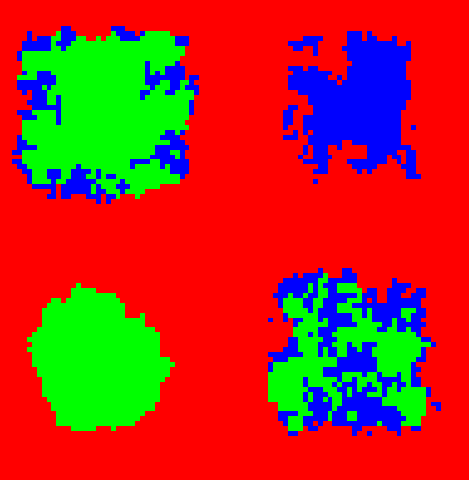} \\
\vspace{0.1cm} 
\hspace{0.05cm}
\includegraphics[width=.42\columnwidth]{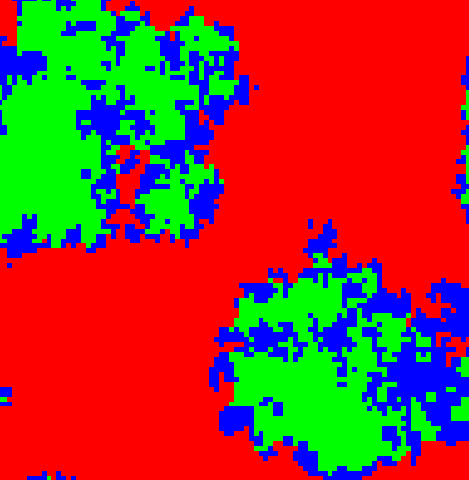}
\includegraphics[width=.42\columnwidth]{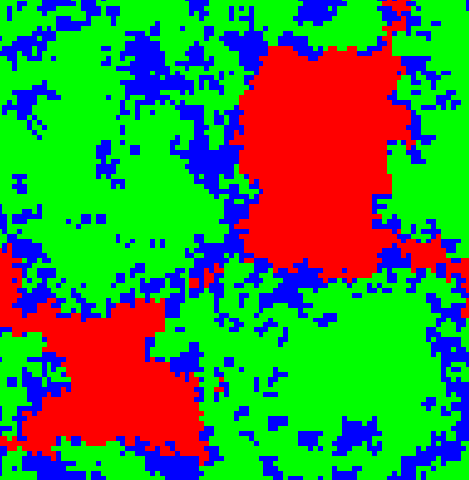}
\caption{{The presence of both altruistic strategies is fundamental to the outcome of a cluster dynamics, allowing the symbiosis configuration to occur.} Here we show the evolution of four different types of clusters in a sea of defectors, with $r = 3.8$, $\gamma=1.5$ {and $K=0.1$} for $t=0$, $40$, $250$ and $500$. For pure cooperator (blue, dark grey) or punisher (green, lightest grey) clusters, the defectors (red, light grey)  quickly dominate. Nevertheless, for both mixed strategy clusters, $C$ and $P$ strategies are able to form spatial structures that both shield from defectors and allow expansion. In ~\cite{gif_fig6} we present a short animation of the lattice evolution.}
\label{fig-time-evo}
\end{center}
\end{figure}

Finally, aiming to confirm the presented micro-{mech\-a\-nism} illustrated in Fig.~\ref{desenho}, we ran simulations with the population prepared in special initial conditions, as presented in Fig.~\ref{fig-time-evo}. 
We considered two squares with pure strategies (top right: cooperators, and bottom left: punishers), and two mixed strategies squares with different configurations (top left and bottom right). In the bottom right, we have a random mix of both strategies, while in the top left we have a bulk of punishers surrounded by a thin layer of cooperators. 

We first note the expected effect on pure populations for the given set of parameters, that is, both pure cooperators and pure punishers quickly disappear.
Nevertheless, we can see that the mixed configurations can survive the {in the} sea of defectors by sticking together. Not only this but they eventually start to expand, dominating the defectors, a feat that neither altruistic strategy can achieve by itself.
As time {progresses}, we note that both types of mixed initial configurations quickly become similar, i.e., a somewhat random distribution of $C$ and $P$, but with the cooperators mostly in the exterior of the cluster.
In this cluster with both strategies present, cooperators can isolate punishers from direct contact with the sea of defectors. 
This avoids high costs for punishers and, on the other hand, if some defectors start to penetrate the cluster, they are quickly defeated by the bulk of punishers, which act as an internal protection. 
We also present a short animation of the lattice evolution in~\cite{gif_fig6}, where it is possible to see the full evolution of this specially prepared initial state and how the mechanism works.

We note again that this complex and interesting mechanism is spontaneous and does not require any external influence on how cooperators and punishers interact.
It is interesting to note how this emergent phenomenon shares many similarities with cyclic alliances {in which} different strategies cyclically dominate most of the population. {This process results} in a system that is more robust to the invasion of foreign strategies ~\cite{Szabo2007,Cazaubiel2017}. Nevertheless, we also note that here the symbiotic effect is achieved without the need of a cyclic dominance structure.

{We extend the analysis by simulating a well-mixed population by allowing agents to copy any other player in the entire population, instead of one of its closest neighbours. Besides this, all players in their public goods groups are also randomly selected in each step. This is done to minimize the effects of the spatial distribution, since this kind of topology is known to neutralize spatial reciprocity.} 

\begin{figure}
\begin{center}
\includegraphics[width=0.9\columnwidth]{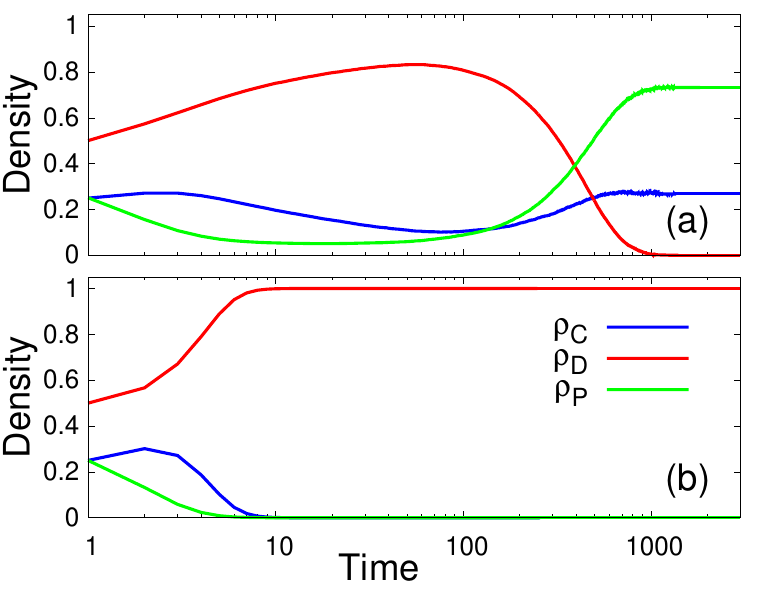}
\caption{Average strategy temporal evolution in the symbiotic parameter region, for $r=3.8$ and $\gamma=1.2$. Panel (a) presents the result for the square lattice, while (b) shows the evolution of the well-mixed population. { The symbiotic effect cannot occur using a random and dynamical spatial structure}, and both cooperators and punishers are dominated by defectors.}
\label{fig-well-mixed}
\end{center}
\end{figure}

Fig.~\ref{fig-well-mixed} presents the population evolution comparing the square lattice with the well-mixed case for the same parameters ($r=3.8$ and $\gamma=1.2$) in the symbiosis region. While altruistic behaviour thrives in the square lattice, it becomes quickly extinct in the well-mixed network. Note that cooperation would also be condemned here if there were no punishers.
We see that even if the parameters are identical, the well-mixed setting forbids the symbiotic phenomenon from taking place due to the lack of spatial organization. This, again, strengthens our claim that the symbiosis seen in our model has a deep relation to the spatial distribution of the strategies.

\section{Summary and Conclusion}

Cooperation is a major component for the integrity of many societies, animal groups, and even inter-species relations. Given the difficulties in maintaining such phenomena, several mechanisms are employed to sustain the altruistic behaviour, such as the punishment of egotistic individuals.
In the current work, we explored the effectiveness of punishment in the public goods game. To do so, altruistic punishers were introduced, whose only condition for punishment to occur is the spatial proximity between the punisher and one or more defectors.
In the classical scenario, without punishment, the survival of cooperation requires an incentive ($ r $) large enough to allow cooperator clusters to avoid invasion by defectors. However, for lower values {of $r$}, cooperators are unable to stand alone, needing an alternative way to fight exploiters. 
With the introduction of punishers, altruistic strategies can survive for a greater range of parameters. 
{For low cost punishments, clusters of punishers are essential for survival, because different agents share the costs of punishing  external defectors while benefiting from spatial reciprocity of internal altruistic behaviour. %
This is in accordance with similar models  }~\cite{brandt2003punishment, Helbing_2010} {where the grouping of punishers was found for appropriate ranges of parameters.}

Nevertheless, an even more interesting phenomenon arises when the punishment value becomes too {high for punishers to thrive alone}. In this parameter region, that we designate the symbiosis region, the fine and $r$ values allow specific spatial configurations to emerge between cooperators and punishers that permit both to survive, whereas none would be able to do so by themselves. 
Such symbiosis achieves stability through spatial patterns that segregate cooperators to borders of the clusters, leaving punishers inside. This allows punishers to survive the costs of high fine values{, and at the same time quickly expel  any defector that invades the cluster.}   { In this manner, } cluster growth  by a mutually beneficial interaction between cooperators and punishers {is possible}.
Analyzing the payoff distribution and lattice snapshots we were able to understand and validate the microscopic mechanism behind this symbiotic formation finding a high sensitivity to the lattice {reciprocity.} 

{The dependence between a cooperator and a punishing strategy to survive was previously found in}~\cite{chen2014probabilistic}. { Differently from our model, they first considered a probabilistic punishment strategy which did not lead to cluster segregation. In other words, both altruistic strategies were randomly distributed at all times.
Furthermore, in the case of non-probabilistic punishers, an altruistic spatial segregation pattern (Fig.~4 from}~\cite{chen2014probabilistic}) {similar to ours emerges; however, it is unstable and vanishes. 
We emphasize that despite the fact that our proposed model does not rely on complex dynamics, such as those of}~\cite{nakamaru2006}, {on extended groups as in}~\cite{brandt2003punishment} {or on diverse punishing rules as in}~\cite{brandt2003punishment,yang2018promoting,chen2014probabilistic,perc2015double, Szolnoki2017a,nakamaru2006},{ it allows a robust survival of both altruistic strategies. 
Moreover,  cooperators are second order free-riders only for a range of parameter values (the V-shaped region presented earlier) since they can only survive if punishers are present to deter defectors as in}~\cite{chen2014probabilistic,Szolnoki2017a,Helbing_2010}. 
{On the other hand, this free-riding situation ceases to exist in the symbiosis region of our work,  since now punishers and cooperators need one another to survive. }

The emergence and maintenance of cooperation is a fascinating subject in both animal evolution and human societies. Specifically, the emergence of symbiotic relations is one of the strongest forms of mutual cooperation in nature, and robust mathematical frameworks to model such associations are a welcome addition to the literature. 
This work opens future venues for studying how symbiosis can spontaneously emerge, and how the connection network can be used to generate 
{configurations with appropriated spatial reciprocity allowing} cooperation to survive in competitive scenarios.

\section*{Acknowledgements}

L.S.F. thanks  FAPERGS (Fundação de {Am\-pa\-ro} à {Pes\-qui\-sa} do Estado do Rio Grande do Sul) for the PROBIC scientific initiation funding and Yan Levin for discussions in the early state of this work. 
H.C.M.F. acknowledges  Universitat de Barcelona where  part  of  this  work was developed. 
M.A.A  was supported by the Brazilian Research Agencies CNPq (proc. 428653/2018-9).
M.H.V. acknowledges his ICTP associateship, where part of this work was developed during a scientific visit to the centre.
The simulations were performed on the IF-UFRGS computing cluster infrastructure.

\section*{Appendix A}
\begin{figure}
\begin{center}
\includegraphics[width=.45\textwidth]{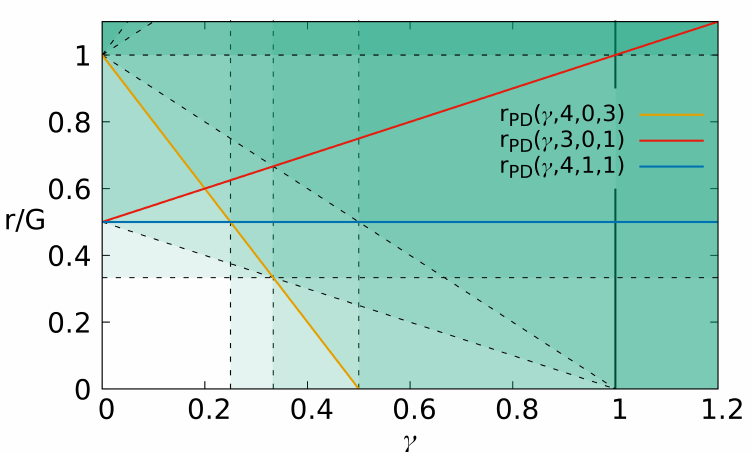}
\caption{Not all transition lines $r_{PD}(\gamma,\NPC+\NPP,\NDC,\NDP)$, Eq.~(\ref{eq.payPD-result}), are decisive for the system's evolution as explained in the text.  Here $\NIJ$  represents the number of sites with strategy  $j$ in central agent $(i)$'s  group, including itself. All non-vertical lines that obey the constraints $\NDC \geq 1 $, $\NCC+\NCP \leq 4$, and $N^{\scriptscriptstyle(i)}_C + N^{\scriptscriptstyle(i)}_D + N^{\scriptscriptstyle(i)}_P=G$ for $i\in\{C,D\}$ are shown.  Likewise, we also present the vertical lines from Eq.~(\ref{eq.vertical-transitions}). Punishers can outcompete defectors by themselves and thrive in the V-shaped region delimited by lines $r_{PD}(\gamma, 4, 0, 3)/G=1-2\gamma$  (yellow) and $r_{PD}(\gamma, 3, 0, 1)/G=(1+\gamma)/2$ (red). The first line corresponds to the change in dominance of a $P-D$ pair on the edge of an altruist cluster; the second corresponds to the same pair in a corner. In the  symbiosis region, delimited by the lines $r_{PD}(\gamma, 4, 1, 1)/G=0.5$ (blue) and the vertical solid line $\gamma_{CD} = 1$, cooperators and punishers can only survive in the presence of one another. Regions above non-vertical lines and to the right of vertical lines favour altruism. We  have  added  a  green  transparency  as  a guide to the eye:  whenever a line is crossed in the  direction  that  favours  punishers  or  cooperators,  the shade becomes darker. }
\label{fig.apend-retas}
\end{center}
\end{figure}

Here we analyse the relevance of the solid line transitions present in Fig.~\ref{fig.par-space}. Figure~\ref{fig.apend-retas} shows possible transition lines arising from all conditions in Eq.~(\ref{eq.payPD-result}) that consider a punisher and a defector as central focal sites, as well as those from Eq.~(\ref{eq.vertical-transitions}) that results from the comparison of the payoffs of a cooperator and a defector as central sites with the constraints given in the main text. These lines  represent the boundaries between  different regions of the system's behaviour under  deterministic evolution, i.e., when $K\to 0$ in Eq.~(\ref{eq.transition}). When $K\neq 0$, there is noise and these lines serve as guides to the whereabouts of the expected transitions.  All $(\gamma,r)$ values  such that  $ r > r_{PD}(\gamma,\NPC+\NPP,\NDC,\NDP)$, for a given set of $N^{(i)}_j$s, favour the punisher in the $P-D$ pair. Here, $\NIJ$  represents the number of sites with strategy  $j$ in central agent $(i)$'s  group, including itself. 
Likewise, regions to the right of each vertical line favour the cooperator in the $C-D$ pair. We have added a green transparency as a guide to the eye in Fig.~\ref{fig.apend-retas}: whenever a line is crossed in the direction that favours punishers or cooperators, the shade becomes darker. 

\begin{figure}[H]
\begin{center}
\includegraphics[width=.18\textwidth]{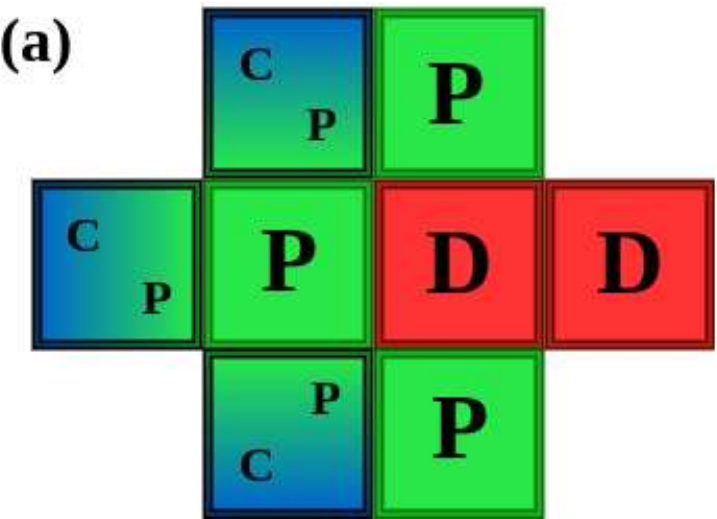}\hspace{0.3cm}
\includegraphics[width=.18\textwidth]{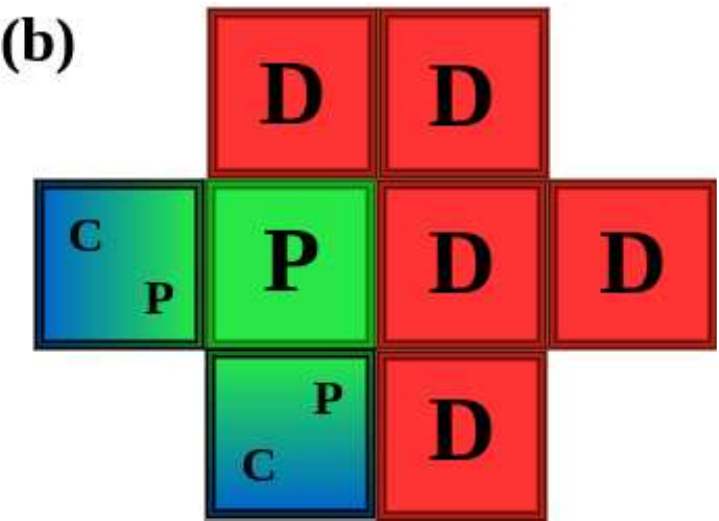} 
\caption{Sites with the letter $P$ or $D$ indicate punishers or defectors, respectively. Sites with letters $C$ and $P$ indicate either a cooperator or a punisher interchangeably. Microscopic configurations of the central pair $P-D$ and their neighbourhoods that govern transitions $r_{PD}(\gamma, 4, 0, 3)/G=1-2\gamma$ and  $r_{PD}(\gamma, 3, 0, 1)/G=(1+\gamma)/2$ are shown in (a) and  (b), respectively. Different outcomes occur whether (a) the central $D$ can invade an altruistic cluster from a border by outcompeting the central $P$ and $(b)$ whether the central $D$ can invade the cluster from a corner. For values of $(\gamma,r)$ in the region above both transition lines $r_{PD}(\gamma, 4, 0, 3)$ and  $r_{PD}(\gamma, 3, 0, 1)$, that delimit the V-shaped region of Fig.~\ref{fig.par-space}, neither process will have a high probability of occurring and the punishers will be favoured. Therefore, in this parameter region, altruistic compact clusters are stable against invasion by defectors.}
\label{fig.append.config.PD}
\end{center}
\end{figure}

There are two main regions of interest to be analysed in Fig.~\ref{fig.apend-retas}. The first one is delimited below by the the solid lines $r_{PD}(\gamma, 4, 0, 3)/G=1-2\gamma$ (yellow line) and $r_{PD}(\gamma, 3, 0, 1)/G = (1+\gamma)/2$ (red line), and corresponds to the V-shaped region of Fig.~\ref{fig.par-space}~(a).  These lines exist whether or not cooperators are present and  delimit the region in which defectors cannot invade compact altruistic clusters composed of $P$s or both $C$s and $P$s. The microscopic configurations corresponding to these boundaries are shown in Fig.~\ref{fig.append.config.PD}. Figure~\ref{fig.append.config.PD}~(a) shows a $P-D$ pair as focal sites and considers a possible invasion of an altruistic cluster from a cluster's border by a defector: if  $r >r_{PD}(\gamma,4,0,3)$, the invasion would not happen in the deterministic case and would have a low probability of succeeding in the presence of noise. In the same manner, Fig.~\ref{fig.append.config.PD}~(b) indicates a possible invasion of an altruistic cluster from a corner when $ r < r_{PD}(\gamma,3,0,1)$. In this way, we conclude that for parameter values outside of this V-shaped region, compact altruistic clusters are unstable.

\begin{figure}[H]

\begin{center}
\includegraphics[width=.36\textwidth]{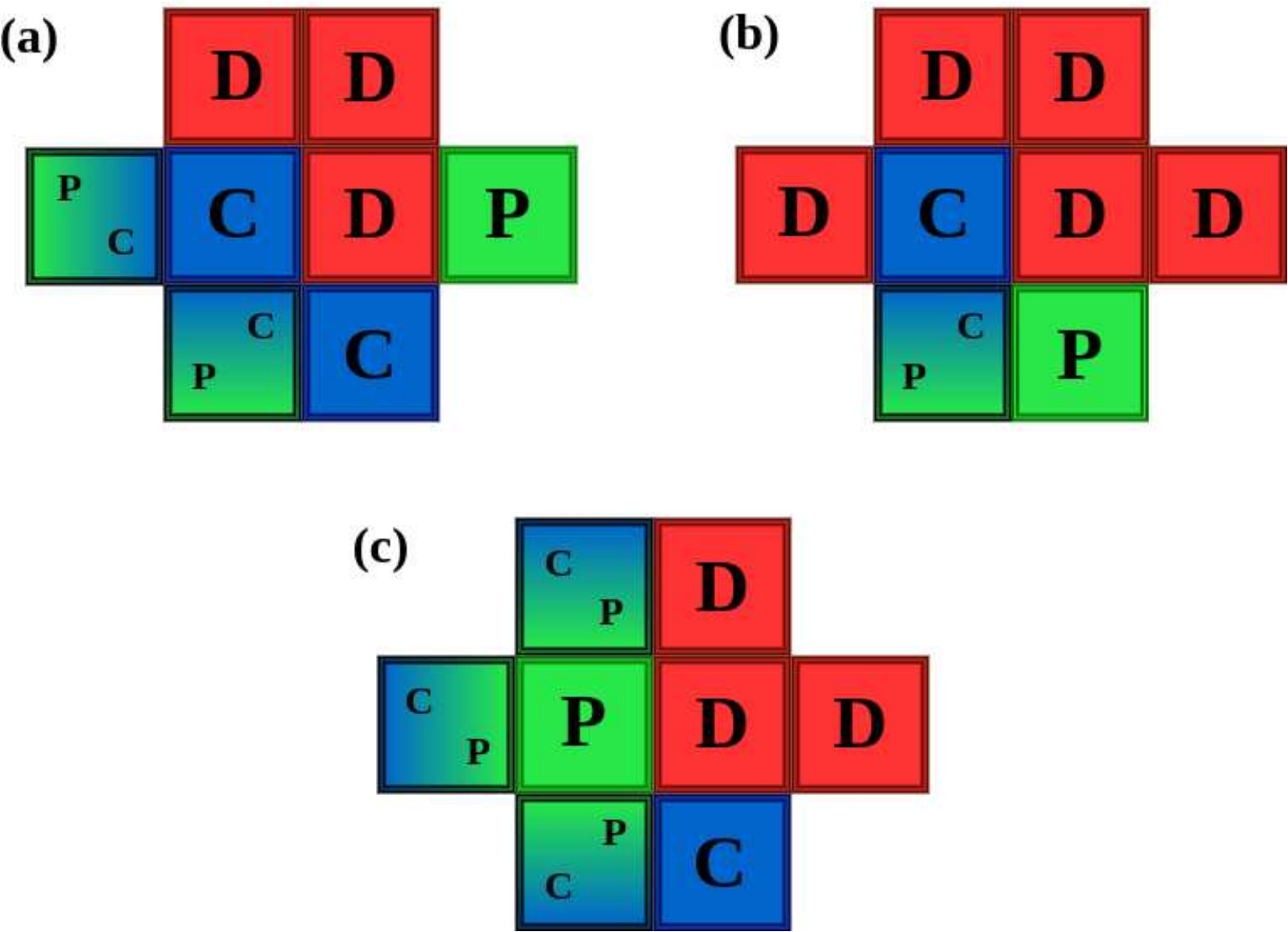} 
\caption{Sites with the letter $C$, $P$ or $D$ indicate cooperators, punishers or defectors, respectively. Sites with letters $C$ and $P$ indicate either a cooperator or a punisher interchangeably. (a) and (b) show microscopic configurations of the central $C-D$ pair and their neighbourhoods responsible for the transition $\gamma_{CD} = 1$, for $\NDP=1$. The constraint $\NDC+\NDP=\NCC+\NCP$ that result in the vertical boundaries equals (a) 3 and (b) 2.   In (c), we present a microscopic configuration of a central $P-D$ pair and their neighbours that govern the transition $r_{PD}(\gamma, 4, 1, 1)/G\equiv 0.5$. When both $\gamma>1$ and $r>r_{PD}(\gamma, 4, 1, 1)$, altruistic clusters are stable and can grow due to the interdependence of cooperators and punishers which characterizes the symbiosis region. }
\label{fig.append.config.CD} 
\end{center}
\end{figure}

However, this scenario changes in the second region of interest, which we denominated the symbiosis region, see Fig.~\ref{fig.par-space}~(c). It only arises in the presence of both cooperators and punishers even though the above conditions for the stability of altruistic  clusters are not met for $\gamma>1$ and  $ r < r_{PD}(\gamma,3,0,1)$. In the deterministic case, in the region to the right of the vertical line $\gamma_{CD} = 1$  and above the solid line $r_{PD}(\gamma, 4, 1, 1)/G\equiv 0.5$ (blue line) in Fig.~\ref{fig.apend-retas},  altruistic clusters once more become stable, grow and dominate the population. In Fig.~\ref{fig.append.config.CD}~(a) and (b) we show microscopic configurations of a focal $C-D$ pair and their neighbourhoods that present a change in behaviour when the vertical line $\gamma_{CD} = 1$ is crossed. For $\gamma < 1$, the central defector has a higher chance of invading the central cooperator, whereas for $\gamma > 1$, it is the cooperator who is more likely to invade the defector. This configuration is shown in Fig.~\ref{fig.append.config.CD}~(b) and Fig.~\ref{desenho} shows how it can be reached. These two configurations indicate that an altruistic cluster can both (a) prevent a defector invasion and  (b) grow into the sea of defectors for $\gamma>1$.  The two configurations occur when $\NDP = 1$ and the constraint $\NCC+\NCP = \NDC+\NDP$ is equal to (a) 3 and (b) 2,  giving rise to the solid vertical line, $\gamma=1$. 
The other condition that delimits the symbiosis region is that the $P$ in the $P-D$ pair shown in Fig.~\ref{fig.append.config.CD}~(c) must have the advantage. For $r> r_{PD}(\gamma, 4, 1, 1)/G\equiv 0.5$, this is indeed the case and an altruistic cluster cannot again be invaded by a defector and grows to dominate the whole population. We emphasize that with noise, or $K\neq 0$, these transitions become diffuse and the above mentioned lines serve only as guides. 
It was by means of these analyses that we were able to identify and better understand which configurations were important in the transition regions. Together with lattice snapshots and the parameter diagram (Fig.~\ref{fig.par-space}) we are able to  understand the microscopic mechanism that leads to the growth and stability of the symbiotic structure composed of punishers and cooperators.

\end{document}